\documentclass[a4paper,11pt]{article}
\usepackage{jheppub}
\usepackage{verbatim}
\usepackage{graphicx}
\usepackage{wrapfig}
\graphicspath{ {./} }
\usepackage{amsmath,verbatim,esint}

\newcommand{\tphi}{\tilde{\phi}}
\newcommand{\tb}{\tilde{b}}

\date{}
\title{Exact Evaluation of Lattice-Regularized Scalar Field Vacuum Amplitude via Site Permutations}
\author{Vadim Asnin}
\affiliation{Independent Researcher}
\emailAdd{asninv@gmail.com}

\abstract{A novel approach to a computation of vacuum amplitude for a single scalar field in any number of dimensions with arbitrary potential on a lattice is proposed. The computation involves symmetrization of kinetic exponential factor in the path integral over all permutations of lattice sites and Waring decomposition of the symmetrized expression.}

\begin{document} 
\maketitle
\section{Introduction}
\label{sec:Introduction}

Path integral is an object of utmost importance in Quantum Mechanics (QM) and Quantum Field Theory (QFT). Whereas in QM problems there are a few alternative approaches (Schrodinger Equation, Heisenberg Equations), in QFT these approaches become too complicated, and the path integral turns out to be a central object to study.  All quantities in QFT are represented through path integrals: correlation functions, scattering amplitudes, vacuum amplitudes etc.

Path Integrals are a central notion in Statistical Mechanics as well. Actually, path integrals of QM and QFT are connected to those of Statistical Mechanics by a Wick rotation of time.

There are basically two well-established general approaches to the computation of path integrals. The first approach is a Perturbation Theory, in which the path integral is expanded in powers of a coupling constant, and each term becomes a calculable Gaussian integral. But in general in QFT most terms diverge at high values of momenta (ultraviolet divergence), and as long as the theory involves massless fields, also at low momenta (infrared divergence). Then some procedure of regularization of Gaussian integrals and a renormalization of coupling constants and fields is required. Another approach, yet intimately connected to the first one, is a Wilsonian Renormalization Group \cite{PhysRevB.4.3174}, which involves a slice-by-slice computation of integrals and also leads to the renormalization of coupling constants and fields. 

In both approaches, the standard computational tool is Gaussian integrals, which are evaluated by means of the Wick theorem. In a language of the Renormalization Group, it is a study of the theory around a Gaussian fixed point. More involved examples of such fixed points are, for instance, a Wilson-Fischer fixed point in dimensional renormalization \cite{PhysRevLett.28.240} and the Banks-Zaks fixed point in supersymmetric large-$N$ QCD \cite{Banks:1981nn}. Once an existence of the fixed point is established one can study perturbations around it.

All the above-mentioned approaches deal with small deviations from known theories. But in many circumstances one needs to analyze significant deviations (strong coupling regimes). Studies of a strong coupling behavior of field theories usually rely on strong-weak dualities ($AdS-CFT$ duality \cite{Aharony_2000}, Seiberg duality \cite{Seiberg_1995} etc), which reduce the problem to the same small deviation paradigm.

An irreparable feature of the Perturbation Theory is a necessity to expand the integrand of the Path Integral in powers of various interaction terms, which reduces rate of convergence in Euclidean signature. For example, in the $\phi^4$ theory in the integrand there appear exponentials of the form $e^{-\frac{m^2}{2}\phi^2-g\phi^4}$. For $g>0$ the integral over $\phi$ converges better than for $g=0$. The perturbative expansion not only puts $g=0$, thus making the convergence worse, but also produces numerous insertions of $\phi^4$ terms in the integrand, further spoiling the convergence. So we would like to avoid this necessity to expand in powers of the interaction to keep the convergence rate as good as possible.

In this paper we propose a novel approach to the computation of a path integral for scalar theory with arbitrary potential on a lattice of any dimension in Euclidean signature. Our approach has no connection to the perturbative expansion. We concentrate on a vacuum amplitude (partition function).

Our general considerations regarding the evaluation of the vacuum amplitude can be summarized as follows:
\begin{itemize}
    \item Vacuum amplitude in the Euclidean signature is given by the following path integral:
    \begin{equation}\label{PathInegralDefinition}
        I_U=\int\;e^{-S[\phi]}\mathcal{D}\phi
        \end{equation}
        Here, the action $S[\phi]$ is given by 
        \begin{equation}
        S[\phi]=\int d^Dx\,\Bigl( \frac 12(\nabla\phi)^2+U(\phi)\Bigr)
        \end{equation}
        $U(\phi)$ is a potential; we assume that it \textit{doesn't involve spacial derivatives of the field}.
    \item We assume that there is an underlying lattice on which the QFT is defined (lattice regularization), and this lattice has a finite amount of sites.
    \item The integral can be rewritten as
    \begin{equation}
        I_U=\int\;S_K[\phi]\,S_P[\phi]\mathcal{D}\phi
    \end{equation}
    Here 
    \begin{equation}
        S_K[\phi]=\exp\Biggl(-\int d^Dx\,\frac12(\nabla\phi)^2\Biggr),\qquad S_P[\phi]=\exp\Biggl(-\int d^Dx\,U(\phi)\Biggr)
    \end{equation}
    \item We can notice that under the assumption that $U$ doesn't involve derivatives the factor $S_P[\phi]$ in the integrand on the path integral is \textit{symmetric under permutations of lattice sites}, since it is a product of independent integrals, one per each site. On the other hand, the factor $S_K[\phi]$ is \textit{not fully symmetric under such permutations}.
    \item A natural idea to evaluate the integral would be to \textit {symmetrize the factor} $S_K[\phi]$ over all sites. It will produce a new factor $S_K^{sym}[\phi]$:
    \begin{equation}
        S_K^{sym}[\phi]=\frac{1}{N!}\sum_PS_K[\phi_P]
    \end{equation}
    The summation here is over all permutations of lattice sites, and $N$ is the number of these sites. $\phi_P$ stands for the field configuration which is obtained from the initial configuration $\phi$ after the permutation $P$. As an illustration of this symmetrization consider a case of $1$-dimensional compactified lattice with $N$ sites. A field configuration $[\phi]$ in this case is just an assignment of a value of the field to each site, we denote these values by $\phi_1,\;\phi_2\ldots\phi_N$. The kinetic exponent $S_K[\phi]$ will look like 
    \begin{multline*}
        S_K[\phi]\equiv S(\phi_1,\ldots\phi_N)=\\=\exp\Biggl(-\frac{(\phi_1-\phi_2)^2}{2}-\frac{(\phi_2-\phi_3)^2}{2}-\ldots-\frac{(\phi_{N-1}-\phi_N)^2}{2}-\frac{(\phi_{N}-\phi_1)^2}{2}\Biggr)    
    \end{multline*} 
    Symmetrization over lattice sites will give the symmetric function
    $$
        S_K^{sym}[\phi]=\frac{1}{N!}\Bigl(S(\phi_1,\phi_2,\phi_3\ldots\phi_N)+S(\phi_2,\phi_5,\phi_8\ldots)+\ldots\Bigl)
    $$
    with all possible permutations of field values. Similar symmetrization can be done for a lattice of any dimensionality.

    Such a symmetrization will transform the path integral
    to the form
    \begin{equation}
        \int\;S_K^{sym}[\phi]\,S_P[\phi]\mathcal{D}\phi
    \end{equation}
    This new integral equals to the previous integral $I_U$. A rigorous proof is based on a standard Group Representation theory for the symmetric group of all permutations of lattice sites.
    \item Since $S_K^{sym}[\phi]$ is a symmetric function under permutations of space points, one may expect that it can be decomposed as
    \begin{equation}\label{Waring}
        S_K^{sym}[\phi]=\sum\limits_{\alpha}\lambda_{\alpha}\,F_{\alpha}(\phi_1)\ldots F_{\alpha}(\phi_N)
    \end{equation}
     This is an infinite-dimensional version of Waring decomposition of higher rank symmetric tensors \cite{Comon2008}. The summation here goes over unspecified index $\alpha$ (it actually can be continuous). Functions $F_{\alpha}(\phi)$ generalize a notion of eigenfunctions of a symmetric function of two variables, so we will call them eigenfunctions of $S_K^{sym}[\phi]$. The numbers $\lambda_{\alpha}$ are eigenvalues corresponding to eigenfunctions $F_{\alpha}(\phi)$. Functions $F_{\alpha}(\phi)$ are assumed to be normalized, otherwise the eigenvalues $\lambda_{\alpha}$ can be absorbed into $F_{\alpha}(\phi)$ by redefining their normalization.
    \item With this expression for $S_K^{sym}[\phi]$ the path integral can be evaluated:
    \begin{multline}\label{ProvisionalPathIntegral}
        I_U=\int\;S_K^{sym}[\phi]\,S_P[\phi]\mathcal{D}\phi
        =\\=\sum\limits_{\alpha}\lambda_{\alpha}\int d\phi_1\ldots d\phi_N\,F_{\alpha}(\phi_1)\ldots F_{\alpha}(\phi_N)e^{-U(\phi_1)-\ldots -U(\phi_N)}=\\=\sum\limits_{\alpha}\lambda_{\alpha}\Bigl(\int d\phi\,F_{\alpha}(\phi)\,e^{-U(\phi)}\Bigr)^N=\sum\limits_{\alpha}\lambda_{\alpha}\,I_{\alpha}^N
    \end{multline}
    \item It should be stressed, however, that the family of functions $F_{\alpha}(\phi)$ is not unique in a sense that there exist numerous representations like eq. (\ref{Waring}). According to the general theory of Waring decomposition, in a finite-dimensional case there exists a minimal representation with smallest possible amount of terms (this amount is called the Waring rank). Infinite-dimensional analog of this statement is unclear.
\end{itemize}
At this point a validity of such an infinite-dimensional generalization of Waring decomposition may look unjustified. In section \ref{sec:BasicReductionOfPI} we give a representation of the path integral in the form of eq. (\ref{ProvisionalPathIntegral}), although obtained in a slightly different way. The summation over $\alpha$ will be a certain double integral, the integrals $I_{\alpha}$ will be expressed as a series of Hermite functions, and the eigenvalues $\lambda_{\alpha}$ will be expressed through a vacuum amplitude of a free field theory on the same lattice, which is assumed to be known, and it is the only place where particular details of the lattice will matter. 

In section \ref{DiscussionSection} we discuss issues regarding the result and its derivation, and also directions for future work.
\section{Explicit Computation of Path Integral by Symmetrization}
\label{sec:BasicReductionOfPI}

We consider QFT of a single scalar field, which is denoted by $\phi$. A vacuum path integral in the Euclidean signature is given by eq. (\ref{PathInegralDefinition}). We assume that the potential $U(\phi)$ does not depend on derivatives of the quantum field. 

Consider such a path integral on a $D$-dimensional lattice $A$ with a step, which we denote as $M^{-1}$, as it has a mass dimension $-1$. A total number of lattice cites will be denoted by $N$. We assume periodic boundary conditions on the lattice to make a notation more clear, but it is not necessary. A discretized action is 
\begin{equation}
S=\frac12\sum\limits_{\vec{k}}\sum\limits_{p=1}^D M^{2-D}(\phi_{m,\vec{k}}-\phi_{m,\vec{k}+\vec{n}_p})^2+\sum\limits_{\vec{k}}U(\phi_{\vec{k}})
\end{equation}
Here $\vec{k}$ is a vector that runs over all lattice sites, $\vec{n}_p$ is a unit lattice vector in direction $p$, and because of the periodic boundary conditions an addition of $\vec{n}_p$ to the sites at the edges brings back to the first site. In the second term on the RHS there is a factor $M^{-D}$, we absorb it into the potential $U$.

Mass dimension of field $\phi$ is $(D-2)/2$; it is convenient to use the mass scale $M$ to make the field dimensionless. We again absorb it into a redefinition of the potential, so, for example, a conventional potential $g\phi^4$ with all the redefinitions will look as $g\,M^{2(D-2)}\phi^4$. 

We are keeping some mass $m_k$ (the subscript $k$ stands for "kinetic") in the kinetic exponential for future reference. In order to simplify the notation in what follows we denote $\mu_k=m_k^2/M^2$. To keep the full action unchanged, we need to shift a mass term in the potential accordingly, namely, subtract the mass term with $\mu_k$ from the potential. We denote this potential with $\mu_k\phi^2/2$ subtracted by $U^{(-\mu_k)}$.

The full discretized path integral is
\begin{multline}
I_U(A)=\int d^N\phi\;\exp\Biggl[-\frac12\sum\limits_{\vec{k}\in A} \sum\limits_{p=1}^D \bigl(\phi_{\vec{k}}-\phi_{\vec{k}+\vec{n}_p}\bigr)^2-\frac{\mu_k}{2}\sum\limits_{\vec{k}\in A}\phi_{\vec{k}}^2\Biggr]\times\\ \times\exp\Biggl[-\sum\limits_{\vec{k}\in A}U^{(-\mu_k)}(\phi_{\vec{k}})\Biggr]
\end{multline}
In the following we will not explicitly specify that the summation over $\vec{k}$ goes over all lattice cites, it will always be the case. 

The second exponential in the integrand (the potential term) is symmetric under any permutation of lattice sites, whereas the  first exponential (the kinetic term) is not fully symmetric. So, the integral will not change if the first term is symmetrized over all possible permutation of lattice sites. In order to use this fact we insert a product of $\delta$-functions and symmetrize them over permutations:
\begin{multline}\label{PathIntegralWithExplicitDelta}
    I_U(A)=\int d^N\phi \;d^N\tphi\;\prod\limits_{i=1}^N\delta(\phi_i-\tphi_i)\;\times\\ \exp\Biggl[-\frac12\sum\limits_{\vec{k}} \sum\limits_{p=1}^D \bigl(\phi_{\vec{k}}-\phi_{\vec{k}+\vec{n}_p}\bigr)^2-\frac{\mu_k}{2}\sum\limits_{\vec{k}}\phi_{\vec{k}}^2\Biggr]\, \exp\Biggl[-\sum\limits_{\vec{k}}U^{(-\mu_k)}(\tphi_{\vec{k}})\Biggr]=\\=
    \frac{1}{N!}\int d^N\phi \;d^N\tphi\;\sum\limits_P \prod\limits_{i=1}^N\delta(\phi_i-\tphi_{P(i)})\;\times \\ \mathrm{exp}\Biggl[-\frac12\sum\limits_{\vec{k}} \sum\limits_{p=1}^D \bigl(\phi_{\vec{k}}-\phi_{\vec{k}+\vec{n}_p}\bigr)^2-\frac{\mu_k}{2}\sum\limits_{\vec{k}}\phi_{\vec{k}}^2\Biggr]\,\mathrm{exp}\Biggl[-\sum\limits_{\vec{k}}U^{(-\mu_k)}(\tphi_{P(\vec{k})})\Biggr]
\end{multline}
Here, a summation over $P$ is a summation over all permutations of fields $\tphi$.

Now we introduce a representation of a symmetrized $\delta$-function, which is used in the subsequent derivation (it is discussed in detail in App. \ref{AppendixSymmetriedDelta}). This representation is
    \begin{multline}\label{DeltaWithRIntegralText}
    \frac{1}{\nu!}\sum\limits_{P}\prod\limits_{i=1}^{\nu}\delta(\xi_{P(i)}-\eta_i)=\\=\int\frac{Dz}{N!}e^{-\sum\limits_r z_r}\int\limits_{(W)}dx\,\prod\limits_{i=1}^{\nu}\Biggl(\sum\limits_{m=0}^{\infty}f_m(\xi_i)z_m\,\tilde{b}_m(x)\Biggr) \prod\limits_{j=1}^{\nu}\Biggl(\sum\limits_{n=0}^{\infty}f_n(\eta_j)\,a_n(x)\Biggr)
\end{multline}
Here $f_n(\xi)$ is a (currently unspecified) discrete complete family of orthonormal functions of the interval $(-\infty,\infty)$. This set will be denoted collectively by $\{f\}$. Objects $a_n(x)$ and $\tilde{b}_n(x)$ are 
\begin{equation}
    a_n(x)=\,p_n^{ix},\qquad \tilde{b}_n(x)=p_n^{-ix},\qquad
    p_0=1,\;p_1=2,\;p_2=3,\;p_4=5\ldots
\end{equation}
Here $p$'s are a unity and all prime numbers. To each object $\tilde{b}_n(x)$ we have attached an integration variable $z_n$. Integral over $x$ is denoted by $(W)$ and, as discussed in App \ref{AppendixSymmetriedDelta}, is in fact a two-dimensional integral:
\begin{equation}
    \int\limits_{(W)}dx \equiv \landupint\limits_{-\infty}^{\infty}\frac{dt}{t^{3/2}}\int\limits_{-\infty}^{\infty}dx\;\kappa(x,t)\qquad
    \kappa(x,t)=\frac12\Bigl(\frac{i}{\pi\,t }\Bigr)^{3/2}e^{-\frac{ix^2}{4t}}
\end{equation}
Here a semicircle on the integral symbol of the $t$-integral shows that the integration path goes above the singularity at 0.

With this identity we can rewrite eq. (\ref{PathIntegralWithExplicitDelta}) as
\begin{multline}\label{FullPathIntegralVsAlgebra0}
    I_U(A)=\int\frac{Dz}{N!}e^{-\sum\limits_r z_r}\int\limits_{(W)}dx\,\int d^N\phi  \exp\Biggl[-\frac12\sum\limits_{\vec{k}} \sum\limits_{p=1}^D \bigl(\phi_{\vec{k}}-\phi_{\vec{k}+\vec{n}_p}\bigr)^2-\frac{\mu_k}{2}\sum\limits_{\vec{k}}\phi_{\vec{k}}^2\Biggr]\times\\ \times\prod\limits_{\vec{q}}\sum\limits_{n=0}^{\infty}f_n(\phi_{\vec{q}})z_n\,\tb_n(x)\,\Bigl[\sum\limits_{n=0}^{\infty}a_n(x)\int d\tphi\;e^{-U^{(-\mu_k)}(\tphi)}\,f_n(\tphi)\Bigr]^N
\end{multline}
We denote by $K^{\{f\}}(A,\mu_k;x)$ the integrated over $z$ kinetic factor of the integral:
\begin{multline}
    K^{\{f\}}(A,\mu_k;x)=\int\frac{Dz}{N!}e^{-\sum\limits_r z_r}\int d^N\phi  \exp\Biggl[-\frac12\sum\limits_{\vec{k}} \sum\limits_{p=1}^D \bigl(\phi_{\vec{k}}-\phi_{\vec{k}+\vec{n}_p}\bigr)^2-\frac{\mu_k}{2}\sum\limits_{\vec{k}}\phi_{\vec{k}}^2\Biggr]\times\\ \times\prod\limits_{\vec{q}}\sum\limits_{n=0}^{\infty}f_n(\phi_{\vec{q}})z_n\,\tb_n(x)
\end{multline}
The subscripts designate explicitly that $K$ depends on our choice of basic functions; this notation will be used in what follows for other functions as well. We rewrite the eq. (\ref{FullPathIntegralVsAlgebra0}) as
\begin{equation}\label{FullPathIntegralVsAlgebra}
I_U(A)=\int\limits_{(W)}dx\, K^{\{f\}}(A,\mu_k;x)\Bigl[\sum\limits_{n=0}^{\infty}a_n(x)\int d\phi\;e^{-U^{(-\mu_k)}(\phi)}\,f_n(\phi)\Bigr]^N
\end{equation}
Instead of computing $K^{\{f\}}(A,\mu_k;x)$ directly we consider a case of a free theory with a mass $m$ after the above-mentioned subtraction, and a vacuum expectation value $v$. This theory has the potential
\begin{equation}
    U_{free}^{(-\mu_k)}(\phi)=\frac{\,m_p^2(\phi-v)^2}{2M^2}
\end{equation}
Here the subscript $p$ stands for "potential". For this theory a value of the path integral is considered to be known, and it is independent of $v$. As regards its dependence on mass, as long as we divided the mass term artificially between kinetic and potential exponents, the whole path integral should depend on $m^2=m_k^2+m_p^2$. To prevent the abuse of notation and to preserve the similarity with $\mu_k$ we denote $\mu_p = m_p^2/M^2$. We denote the free path integral as $I_{free}\bigl(A;\mu_k+\mu_p\bigr)$. We have
\begin{multline}\label{IntegralEqulation}
     I_{free}\Bigl(A;\mu_k+\mu_p\Bigr)=\int\limits_{(W)}dx\,K^{\{f\}}(A,\mu_k;x)\Bigl[\sum\limits_{n=0}^{\infty}a_n(x)\int d\phi\;e^{-\frac{\mu_p(\phi-v)^2}{2}}\,f_n(\phi)\Bigr]^N
\end{multline}
We would like to consider the eq. (\ref{IntegralEqulation}) as an integral equation for the unknown function $K^{\{f\}}(A,\mu_k;x)$. A kernel of this integral equation is
\begin{equation}\label{IntegralEqKernel}
    P^{\{f\}}(\mu_p,v;x)= \Bigl[\sum\limits_{n=0}^{\infty}\,a_n(x)\int d\phi\;e^{-\frac{\mu_p(\phi-v)^2}{2}}f_n(\phi)\Bigr]^N
\end{equation}
In the integral equation (\ref{IntegralEqulation}) this kernel acts on $K^{\{f\}}(A,\mu_k;x)$ considered as a function of $x$, which results in the function of $\mu_p$. 
In order to solve the integral equation (\ref{IntegralEqulation}) we need to define and compute the inverse kernel $Q^{\{f\}}(\mu_p;y)$, which would satisfy the following integral equation:
\begin{equation}\label{InverseKernelEquation}
    \int d\mu_p\,Q^{\{f\}}(\mu_p;y)\, P^{\{f\}}\Bigl(\mu_p,v(\mu_p);x\Bigr) =\delta^{(a,b)}(x,y)
\end{equation}
Here $\delta^{(a,b)}$ is a distribution that behaves similarly to the Dirac $\delta$-function under $\int^{(a,b)}$ integration. It is defined in App. \ref{AppendixSymmetriedDelta}, eq. (\ref{DeltaABDefinition}); it is shown in eq. (\ref{ZetaDelta}) that this function is actually the Riemann $\zeta$-function $\delta^{(a,b)}(x,y)=\zeta(-i\,(x-y))$. We will think of $v$ as some function of $\mu_p$, a particular form of this dependence will be fixed later; in what follows we will continue to write $v$ and not $v(\mu_p)$. The integral over $\mu_p$ is assumed to be taken from $0$ to $\infty$, the integration limits will not be stated explicitly to avoid an abuse of notation. 

With the inverse kernel $Q^{\{f\}}(\mu_p;y)$ we can write the solution of the integral equation (\ref{IntegralEqulation}) as
\begin{equation}\label{EquationForE}
    K^{\{f\}}\Bigl(A,\mu_k;x\Bigr)= \int d\mu_p\,I_{free}\Bigl(A;\mu_k+\mu_p\Bigr)\,Q^{\{f\}}(\mu_p;x) 
\end{equation}
Rewrite eq. (\ref{InverseKernelEquation}) as
\begin{equation}\label{InverseKernelEquation1}
    \int d\mu_p\,Q^{\{f\}}(\mu_p;y) \Biggl[\sum\limits_{n=0}^{\infty}a_n(x)\int d\phi\;e^{-\frac{\mu_p}{2}(\phi-v)^2}\,f_n(\phi)\Biggr]^N=\delta^{(a,b)}(x,y)
\end{equation}
In order to proceed, we introduce an expansion of the exponential $e^{-\frac{\mu_p}{2}(\phi-v)^2}$ into basic functions $f_n(\phi)$:
\begin{equation}\label{DefinitionOfC}
    e^{-\frac{\mu_p}{2}(\phi-v)^2}=\sum\limits_{q=0}^{\infty} C_q^{\{f\}}(\mu_p,v)\,f_q(\phi)
\end{equation}
Then the expression inside the brackets in eq. (\ref{InverseKernelEquation1}) becomes
\begin{equation}
    \sum\limits_{n=0}^{\infty}a_n(x)\int d\phi\;e^{-\frac{\mu_p(\phi-v)^2}{2}}\,f_n(\phi)=\sum\limits_{n=0}^{\infty}C_n^{\{f\}}(\mu_p,v)\,a_n(x)
\end{equation}
We can rewrite the eq. (\ref{InverseKernelEquation1})
\begin{multline}\label{InverseKernelEquation2}
    \delta^{(a,b)}(x,y)=\int d\mu_p\,Q^{\{f\}}(\mu_p;y)\Biggl[\sum\limits_{n=0}^{\infty} C_n^{\{f\}}(\mu_p,v)\,a_n(x)\,\Biggr]^N\equiv
    \\
    \equiv\sum\limits_{n_1\geq n_2\ldots \geq n_N=0}^{\infty}
    \begin{pmatrix} N\\ n_1,\ldots n_N \end{pmatrix}\,
    a_{n_1}(x)\ldots a_{n_N}(x)\times
    \\
    \times\int d\mu_p\,Q^{\{f\}}(\mu_p;y)\,C_{n_1}^{\{f\}}(\mu_p,v)\ldots C_{n_ N}^{\{f\}}(\mu_p,v)
\end{multline}
Here $\begin{pmatrix} N\\ n_1,\ldots n_N \end{pmatrix}$ are  multinomial coefficients.

We are going to show that in order to solve eq. (\ref{InverseKernelEquation2}) in the limit of large $N$  $Q^{\{f\}}(\mu_p;y)$  should satisfy
\begin{multline}\label{QRequirement}
     \int d\mu_p\,Q^{\{f\}}(\mu_p;y)\,C_{n_1}^{\{f\}}(\mu_p,v)\ldots C_{n_ N}^{\{f\}}(\mu_p,v)= \\=\begin{pmatrix} N\\ n_1,\ldots n_N \end{pmatrix}^{-1}
     \tb_{n_1}(y)\ldots \tb_{n_N}(y)
\end{multline}
In order to prove the correctness of this statement, we substitute it into eq. (\ref{InverseKernelEquation2}) and get
\begin{equation}
    \sum\limits_{n_1\geq n_2\geq\ldots\geq n_N=0}^{\infty}
     \tb_{n_1}(y)\ldots \tb_{n_N}(y)\,a_{n_1}(x)\ldots a_{n_N}(x)=\delta^{(a,b)}(x,y)
\end{equation}
With the explicit construction of $a$'s and $\tb$'s described in App. \ref{AppendixSymmetriedDelta} the LHS of the last equation becomes
\begin{equation}
    \sum\limits_{n_1\geq n_2\geq\ldots\geq n_N=0}^{\infty} \Bigl(p_{n_1}\ldots p_{n_N}\Bigr)^{i(x-y)}=\sum\limits_L L^{i(x-y)}
\end{equation}
with $L\equiv p_{n_1}\ldots p_{n_N}$ being a product of up to $N$ prime numbers. As $N$ becomes very large, this sum eventually turns into a sum over all natural numbers, so it will become
\begin{equation}
   \sum\limits_{L=1}^{\infty}L^{i(x-y)}=\zeta\bigl(-i(x,y)\bigr)
\end{equation}
As mentioned above, this function is actually $\delta^{(a,b)}(x,y)$, so our requirement (\ref{QRequirement}) indeed provides a solution to the eq. (\ref{InverseKernelEquation2}), at least for large values of $N$, and this is the case that we need.

Eq. (\ref{QRequirement}) describes the action of the kernel $Q^{\{f\}}(\mu_p;y)$ on various products of $N$ coefficients $C_n^{\{f\}}(\mu_p,v)$. Instead of considering this action directly, we turn to the definition of $C_n^{\{f\}}(\mu_p,v)$ given in eq. (\ref{DefinitionOfC}). From this equation it follows that
\begin{equation}
    e^{-N\frac{\mu_p(\phi-v)^2}{2}}=\Biggl[\sum\limits_{q=0}^{\infty} C_q^{\{f\}}(\mu_p,v)\,f_q(\phi)\Biggr]^N
\end{equation}
According to eq. (\ref{QRequirement}) we should have
\begin{multline}\label{QRequirement2}
    \int d\mu_p\,Q^{\{f\}}(\mu_p;y)\,e^{-\frac N2 \mu_p\bigl(\phi-v\bigr)^2}=\\=\sum\limits_{n_1\geq n_2\geq\ldots\geq n_N=0}^{\infty}f_{n_1}(\phi)\ldots f_{n_N}(\phi)\,\tb_{n_1}(y)\ldots \tb_{n_N}(y)
\end{multline}
After substituting the explicit form of $\tb_n$'s and taking $N$ to be very large, the RHS of the last equation acquires a form of a Dirichlet series:
\begin{multline}\label{DefinitionOfT}
    \sum\limits_{n_1\geq n_2\geq\ldots\geq n_N=0}^{\infty}f_{n_1}(\phi)\ldots f_{n_N}(\phi)\,\tb_{n_1}(y)\ldots \tb_{n_N}(y)=
    \\
    =\sum\limits_{n_1\geq n_2\geq\ldots\geq n_N=0}^{\infty}f_{n_1}(\phi)\ldots f_{n_N}(\phi)\,p_{n_1}^{-iy}\ldots p_{n_N}^{-iy}
    \sim
    \sum\limits_{r=0}^{\infty}T_r^{\{f\}}(\phi)r^{-iy}
\end{multline}
Here $T_r^{\{f\}}(\phi)$ is defined as follows: if $r$ is decomposed into primes as $r=p_1^{n_1}\ldots p_k^{n_k}$ (according to the definitions of App. \ref{AppendixSymmetriedDelta} $p_0=1$, we define $n_0=N-n_1-\ldots n_k$) then 
\begin{equation}
    T_r^{\{f\}}(\phi)=f_0(\phi)^{n_0}f_1(\phi)^{n_1}\ldots f_k(\phi)^{n_k}
\end{equation}
In what follows we will denote for short
\begin{equation}    
    T^{\{f\}}(\phi,y)=\sum\limits_{r=0}^{\infty}T_r^{\{f\}}(\phi)r^{-y}
\end{equation}
Quantities $T_r^{\{f\}}(\phi)$ constitute a completely multiplicative sequence, therefore $T^{\{f\}}(\phi,y)$ has in fact a form of the Euler product in the limit of large $N$
\begin{equation}
    T^{\{f\}}(\phi,y)\sim f_0(\phi)^N\prod\limits_{n=1}^{\infty}\frac{1}{1-\frac{f_n(\phi)}{f_0(\phi)}\,p_n^{-y}}
\end{equation}
With these definitions eq. (\ref{QRequirement2}) becomes
\begin{equation}
\int\limits_0^{\infty} d\mu_p\,Q^{\{f\}}(\mu_p;y)\,e^{-\frac N2 \mu_p\Bigl(\phi-v(\mu_p)\Bigr)^2}=T^{\{f\}}(\phi,iy)
\end{equation}
Here we have explicitly written the dependence of $v$ on $\mu_p$. We choose the function $v(\mu_p)$ to be linear: $v(\mu_p)=a\mu_p$ with $a>0$. Make a change of the integration variable $\mu_p\rightarrow \mu_p/N$, it gives
\begin{equation}
\int\limits_0^{\infty} d\mu_p\,Q^{\{f\}}\biggl(\frac{\mu_p}{N};y\biggr)\,e^{-\frac{\mu_p}{2}\Bigl(\phi-a\mu_p/N\Bigr)^2}=N\,T^{\{f\}}(\phi,iy)
\end{equation}
In the limit of large $N$ we can omit the $1/N$ term in the exponent and get
\begin{equation}
\int\limits_0^{\infty} d\mu_p\,Q^{\{f\}}\biggl(\frac{\mu_p}{N};y\biggr)\,e^{-\frac{\mu_p}{2}\phi^2}=N\,T^{\{f\}}(\phi,iy)
\end{equation}
This means that $Q^{\{f\}}$ is basically a Laplace transform of $T^{\{f\}}$, the solution is given by the inversion formula
\begin{equation}
    Q^{\{f\}}\bigl(\mu_p;y\bigr)=\frac{N}{2\pi i}\int\limits_{\Gamma} T^{\{f\}}(v,iy)\,e^{N\mu_pv^2/2}v\,dv
\end{equation}
Here $\Gamma$ stands for the Bromwich line, its place on the real line will be discussed later, together with a convergence of the integral. Eq. (\ref{EquationForE}) gives the following expression for the kinetic factor:
\begin{equation}
    K^{\{f\}}\Bigl(A,\mu_k;x\Bigr)= \frac{N}{2\pi i}\int\limits_0^{\infty}d\mu_p\int\limits_{\Gamma}\,I_{free}\Bigl(A;\mu_k+\mu_p\Bigr)\,T^{\{f\}}(v,ix)\,e^{N\mu_pv^2/2}v\,dv 
\end{equation}
With the last expression for $K^{\{f\}}\Bigl(A,\mu_k;x\Bigr)$ we can rewrite the full path integral (see eq. (\ref{FullPathIntegralVsAlgebra})) as
\begin{multline}\label{FullResult}
    I_U(A)=\frac{N}{2\pi i}\int\limits_0^{\infty}d\mu_p\int\limits_{\Gamma}dv \int\limits_{(W)}dx\,I_{free}\Bigl(A;\mu_k+\mu_p\Bigr) T^{\{f\}}(v,ix)\,e^{N\mu_pv^2/2}v\times 
     \\
     \times\Bigl[\sum\limits_{n=0}^{\infty}\,p_n^{ix}\int d\phi\;e^{-U^{(-\mu_k)}(\phi)}\,f_n(\phi)\Bigr]^N
\end{multline}
At this point we finally choose our basic functions to be scaled Hermite functions
\begin{equation}\label{ScaledHermiteBasicFunctions}
    f_n(\phi)=\sqrt{\mu}\,\psi_n(\sqrt{\mu}\,\phi)
\end{equation}
With this choice our result becomes
\begin{multline}\label{FullResultHermite}
    I_U(A)=\frac{N}{2\pi i}\int\limits_0^{\infty}d\mu_p\int\limits_{\Gamma}dv \int\limits_{(W)}dx\,I_{free}\Bigl(A;\mu_k+\mu_p\Bigr) T^{\{f\}}(v,ix)\,e^{N\mu_pv^2/2}v\times 
     \\
     \times\Bigl[\sqrt{\mu}\sum\limits_{n=0}^{\infty}\,p_n^{ix}\int d\phi\;e^{-U^{(-\mu_k)}(\phi)}\,\psi_n(\sqrt{\mu}\,\phi)\Bigr]^N
\end{multline}
As a simple check of eq. (\ref{FullResultHermite}) consider the case of a free theory with the potential
\begin{equation}
    U^{(-\mu_k)}(\phi)=\frac{\mu}{2}\phi^2
\end{equation}
For this potential we have 
\begin{equation}
    e^{-U^{(-\mu_k)}(\phi)}=\frac{\pi^{1/4}}{\sqrt{\mu}}f_0(\phi)
\end{equation}
Then it follows
\begin{equation}
    \sum\limits_{n=0}^{\infty}\,p_n^{ix}\int d\phi\;e^{-U(\phi)}\,f_n(\phi)=\frac{\pi^{1/4}}{\sqrt{\mu}}
\end{equation}
This integral is independent of $x$, therefore, using eq. (\ref{DefinitionOfT})
\begin{multline}
    \int\limits_{(W)}dx\,\int\limits_{\Gamma} dv\,v\,T^{\{f\}}(v,ix) e^{N\mu_pv^2/2}\Bigl[\sum\limits_{n=0}^{\infty}\,p_n^{ix}\int d\phi\;e^{-U^{(-\mu_k)}(\phi)}\,f_n(\phi)\Bigr]^N=\\=\Bigl(\frac{\pi}{\mu^2}\Bigr)^{N/4}\int \limits_{\Gamma}dv\,v\,e^{N\mu_pv^2/2}\,f_0(v)^N=\int\limits_{\Gamma} dv\,v\,e^{N(\mu_p-\mu) v^2/2}=\\=\frac{2\pi i}{N}\,\delta(\mu_p-\mu)
\end{multline}
The integral over $v$ goes along the straight (Bromwich) line, so the integral over $v^2$ goes over a curved line, which, however, can be deformed to become straight. We see that we can place the Bromwich line at any point at the positive semi-axis. Then, for this free theory we get as it should be
\begin{equation}
    \frac{N}{2\pi i}\int\limits_0^{\infty} d\mu_p\,I_{free}\Bigl(A;\mu_k+\mu_p\Bigr)\frac{2\pi i}{N}\delta(\mu_p-\mu)\equiv I_{free}\Bigl(A;\mu_k+\mu\Bigr)
\end{equation}

A few remarks about the eq. (\ref{FullResultHermite}) are in order. This is the expression of the form of eq. (\ref{ProvisionalPathIntegral}). We see the following concrete form of various components of eq. (\ref{ProvisionalPathIntegral}):
\begin{itemize}
    \item The summation over $\alpha$ gets replaced by the $(W)$-integration .
    \item Eigenfunctions $F_{\alpha}(\phi)$ got replaced by
    $$F_{\alpha}(\phi)\rightarrow F_x(\phi)=\sqrt{\mu}\sum\limits_{n=0}^{\infty}\psi_n(\sqrt{\mu}\,\phi)\,p_n^{ix}
    $$
    This series is not convergent, so we cannot use these functions directly. However, the series
    \begin{equation}
        I_n(x)=\sqrt{\mu}\sum\limits_{n=0}^{\infty}\,p_n^{ix}\int d\phi\;e^{-U^{(-\mu_k)}(\phi)}\,\psi_n(\sqrt{\mu}\,\phi)
    \end{equation}
    is convergent, since the integrals are coefficients of an expansion of $e^{-U^{(-\mu_k)}(\phi)}$ into Hermite functions, and therefore decrease rapidly with $n$.
    \item Eigenvalues $\lambda_{\alpha}$ have become $$\lambda_{\alpha}\rightarrow\lambda_x=\frac{N}{2\pi i}\int\limits_0^{\infty}d\mu_p\int\limits_{\Gamma}dv\, I_{free}\Bigl(A;\mu_k+\mu_p\Bigr) T^{\{f\}}(v,ix)\,e^{N\mu_pv^2/2}v$$
    
    \item We haven't found an expansion of the symmetrized kinetic exponential into its eigenfunctions, nevertheless our expression for the path integral is of the form of eq. (\ref{ProvisionalPathIntegral}), and this is what we needed.
    \item The parameter $\mu_k$ and the mass squared scale $\mu$ can be chosen arbitrarily. 
    \item We have established that eq. (\ref{FullResultHermite}) correctly reproduces path integral for free theory.
    
\end{itemize}

\section{Discussion}\label{DiscussionSection}

In this paper we described a method of computing a vacuum amplitude of single scalar quantum field on a lattice in the Euclidean signature. This method is applicable under the assumption that the QFT action can be separated into a standard kinetic term and a potential term, which doesn't contain spatial derivatives of the field. This method consists of the following logical steps: 1) Symmetrization of kinetic exponential w.r.t all permutations of lattice sites; 2) Separation of fields in the kinetic and potential exponentials by inserting the appropriate representation of symmetrized $\delta$-function of fields; this step directly leads to the representation of the results in the form of eq. (\ref{ProvisionalPathIntegral}); 3) Computation of eigenvalues by comparison to a free theory. A few remarks about this approach are in order. 
\begin{enumerate}
    \item We considered in detail the case of a single scalar field. Our approach can be generalized to a case of more scalar fields, provided that the potential doesn't include their derivatives. Indeed, the main technical tool, which is the identity (\ref{DeltaWithRIntegral}), can be used also for this case, as long as basic functions satisfy eq. (\ref{BasicCompletenessRelation}), where now $\xi$ and $\eta$ are points in $k$-dimensional space, where $k$ in a number of quantum fields.
    \item We have not dealt with neither spinor nor calibration fields. A reason for that, except for technical complications, is that theories like QED or non-Abelian gauge theory have terms with derivatives of fields in the interaction Lagrangian. However, our approach can in principle be applied to Yukawa theory or four-fermion interaction.
    \item One can use our result (\ref{FullResultHermite}) as a starting point of a derivation of Wilsonian RG flow.
    \item We did not address the problem of computing correlation functions of fields. In order to compute a correlation function, which is represented by a path integral similar to the one considered in this work, but with insertions of fields, one can in principle follow the same route: symmetrize the part of the integrand which is not symmetric under permutation of lattice sites, namely, the kinetic exponential together with field insertions.
\end{enumerate}

\section*{Acknowledgements}
The author is grateful to Dr. P. Renkel for lengthy discussions at various stages of this work.

\appendix
\section{Identity for symmetrized $\delta$-function} \label{AppendixSymmetriedDelta}
Consider a complete set of functions $f_n(\xi)$, $n=0\ldots\infty$ on the interval $(-\infty,\infty)$. Such a family of functions satisfies the completeness identity
\begin{equation}\label{BasicCompletenessRelation}
    \sum\limits_{n=0}^{\infty}f_n(\xi)f_n(y)=\delta(\xi-\eta)
\end{equation}
Consider a product on $\nu$ such identities for variables $\xi_i$, $\eta_i$ with $i=1\ldots \nu$. It is
\begin{equation}\label{MultidimDelta}
    \prod\limits_{i=1}^{\nu} \Bigl(\sum\limits_{n=0}^{\infty}f_n(\xi_i)f_n(\eta_i)\Bigr)= \prod\limits_{i=1}^{\nu}\delta(\xi_i-\eta_i)
\end{equation}
Our purpose is to symmetrize this last equation w.r.t. all permutations of $\xi$'s. The symmetrized identity is 
\begin{equation}\label{SymmetrizedDelta}
    \frac{1}{\nu!}\sum\limits_{P} \prod\limits_{i=1}^{\nu} \Bigl(\sum\limits_{n=0}^{\infty}f_n(\xi_{P(i)})f_n(\eta_i)\Bigr)= \frac{1}{\nu!}\sum\limits_{P}\prod\limits_{i=1}^{\nu}\delta(\xi_{P(i)}-\eta_i)
\end{equation}
Here, $P$'s are permutations of indices $1\ldots \nu$.

Consider a LHS of eq. (\ref{SymmetrizedDelta}) in detail. If we multiply all the parentheses we will have on the LHS a sum of terms of the form $$f_{n_1}\bigl(\xi_{P(1)}\bigr)\ldots f_{n_{\nu}}\bigl(\xi_{P(\nu)}\bigr)\,f_{n_1}(\eta_1)\ldots f_{n_{\nu}}(\eta_\nu)$$ Count how many times each such term appears in the summation. In eq. (\ref{MultidimDelta}) there appear only terms of the form $$f_{n_1}\bigl(\xi_1\bigr)\ldots f_{n_{\nu}}\bigl(\xi_{\nu}\bigr)\,f_{n_1}(\eta_1)\ldots f_{n_{\nu}}(\eta_{\nu})$$ Coefficients of these terms in  (\ref{MultidimDelta}) are all equal to 1. In the symmetrized expression terms which have common values of indices get mixed. After the symmetrization, each term appears $\nu_1!\ldots \nu_k!$ times. Therefore, in eq. (\ref{SymmetrizedDelta}) each term on the LHS appears with coefficient $$ \begin{pmatrix} \nu\\ \nu_1\;\nu_2\ldots \nu_k \end{pmatrix}^{-1} =\frac{\nu_1!\ldots \nu_k!}{\nu!}$$ Here $\begin{pmatrix} \nu\\ \nu_1\;\nu_2\ldots \nu_k \end{pmatrix}$ is a multinomial coefficient.

We now need to introduce new objects, call them $a_n$ and $b_n$, $n=0,1\ldots$ with the following properties:
\begin{itemize}
    \item They form a commutative algebra (so that they can be multiplied by ordinary numbers, can be added and multiplied by each other). We will call this algebra the $(a,b)$-algebra.
    \item There exists an inner product that we denote by $\Bigl<\ldots,\ldots\Bigr>$ such that $\Bigl<b_0^{n_0}b_1^{n_1}\ldots,a_0^{m_0}a_1^{m_1}\ldots\Bigr>=\begin{pmatrix} \nu\\ \nu_1\;\nu_2\ldots \nu_k \end{pmatrix}^{-1}$ if the sets of indices $m$ and $n$ coincide, and 0 otherwise. 
\end{itemize}
Using these objects, we can rewrite the eq. (\ref{SymmetrizedDelta}) as
\begin{equation}\label{DeltaWithAB}
    \Bigl< \prod\limits_{i=1}^{\nu}\sum\limits_{n=0}^{\infty}f_n(\xi_i)b_n, \,\prod\limits_{i=1}^{\nu}\sum\limits_{n=0}^{\infty}f_n(\eta_i)a_n\,\Bigr>=\frac{1}{\nu!}\sum\limits_{P}\prod\limits_{i=1}^{\nu}\delta(\xi_{P(i)}-\eta_i)
\end{equation}
Different realizations of the $(a,b)$-algebra lead to different forms of the last equation. We now build a realization $R$ of the $(a,b)$ algebra which is used throughout the paper. In fact, we build a family of realizations. 
\begin{itemize}
   \item The construction involves the following sequence:
   \begin{equation}\label{DefinitionOfPrimes}
       p_0=1,\; p_1=2,\; p_2=3,\;p_3=5,\;p_4=7,\ldots
   \end{equation}
   which is a unity and all prime numbers.
    \item The algebra elements $a_n$ and $b_n$ are realized as follows (we introduce additional objects $\tb_n$ which will also be useful):
    \begin{equation}\label{AsBs}
        a_n=p_n^{ix},\qquad \tb_n=p_n^{-ix}, \qquad b_n=z_n\,\tb_n
    \end{equation}
    Here $x$ is a variable, and $z_n$ are auxiliary variables needed to fix the normalization.
    \item A product of a few $a$'s will have the form $r^{ix}$, with $r$ being a product of primes that appear in these $a$'s; similarly, a product of a few $b$'s will be of the form $Z_s\,s^{-ix}$, where now $s$ is a product of primes that make up $b$, and $Z_s$ is a product of its $z$ factors. Each set of primes gives its own product, different sets of primes give different products.
    \item We establish a inner product of $a$'s and $\tb$'s as
    \begin{equation}\label{AlgebraIntegral}
        \Bigl<\tb_{n_1}\ldots \tb_{n_{\nu}},a_{m_1}\ldots a_{m_{\nu}}\Bigr>=\frac12\Bigl(\frac{i}{\pi }\Bigr)^{3/2}\landupint\limits_{-\infty}^{\infty}\frac{dt}{t^{3/2}}\int\limits_{-\infty}^{\infty}dx\,e^{-\frac{ i\,x^2}{4t}}\Bigl(\frac rs\Bigr)^{ix}
    \end{equation}
 Here, the contour symbol means that the integral over $t$ circumvents zero from above. Throughout the paper we will denote the joint $(x,t)$-integral in (\ref{AlgebraIntegral}) as
 \begin{equation}
     \int\limits_{(W)}dx
 \end{equation} 
Here $W$ stands for Waring. We denote by $\kappa(x,t)$ the integration kernel
 \begin{equation}
     \kappa(x,t)=\frac12\Bigl(\frac{i}{\pi\,t }\Bigr)^{3/2}e^{-\frac{ix^2}{4t}}
 \end{equation}In order to show that this expression possess the required property we rewrite it as
    \begin{equation}
        \Bigl<\tb_{n_1}\ldots \tb_{n_{\nu}},a_{m_1}\ldots a_{m_{\nu}}\Bigr>= \int\limits_{(W)}dx\,e^{i\alpha\,x}
    \end{equation}
    Here $\alpha=\ln{r}-\ln{s}$. Integration over $x$ leads to
    \begin{equation}
        \Bigl<\tb_{n_1}\ldots \tb_{n_{\nu}},a_{m_1}\ldots a_{m_{\nu}}\Bigr>=  \landupint\limits_{-\infty}^{\infty}\frac{i\,dt}{\pi t}e^{i\,\alpha^2 t}
    \end{equation}
    We decompose the integral to the principal value and the integral over infinitesimal semicircle:
    \begin{equation}
        \landupint\limits_{-\infty}^{\infty}=\fint\limits_{-\infty}^{\infty}+\landupint
    \end{equation}
    The integral over semicircle is
    \begin{equation}
        \landupint\frac{i\,dt}{\pi t}e^{i\,\alpha^2 t}=1
    \end{equation}
    In order to compute the principal value consider the cases of $\alpha=0$ and $\alpha\neq 0$. If $\alpha=0$ then
    \begin{equation}
        \fint\limits_{-\infty}^{\infty}\frac{i\,dt}{\pi t}=0
    \end{equation}
    If $\alpha\neq 0$ then
    \begin{equation}
        \fint\limits_{-\infty}^{\infty}\frac{i\,dt}{\pi t}e^{i\,\alpha^2 t}=-\fint\limits_{-\infty}^{\infty}\frac{dt}{\pi t}\sin{\alpha^2t}=-1
    \end{equation}
    So, the whole integral is $1$ if $\alpha=0$ and $0$ otherwise, as required. We have established that
    \begin{equation}\label{AlgebraIntegralCases}
        \int\limits_{(W)}dx\,e^{i\alpha\,x} = \begin{cases}
    1, & \text{if $\alpha=0$}.\\
    0, & \text{otherwise}.
  \end{cases}
    \end{equation}
    If $\alpha=0$ then all normalization factors $w$ cancel out. Therefore the full inner product is $1$ if the sets of indices $m$ and $n$ coincide, and $0$ if they don't.
    \item Now we can define the inner product of $a$'s and $b$'s. We define it to be 
     \begin{equation}\label{AlgebraIntegralFull}
        \Bigl<b_{n_1}\ldots b_{n_{\nu}},a_{m_1}\ldots a_{m_{\nu}}\Bigr>=\int\limits_{(W)} dx\,\Bigl(\frac rs\Bigr)^{ix}\,\int\limits_0^{\infty}\frac{Dz}{\nu!}\,Z_s\,e^{-z_0-z_1-\ldots}
    \end{equation}
    Here, as before, $Z_s$ is the product of all $z$-factors that appear in the definition of $b$'s. The last integral in this expression is formally infinitely dimensional (it consists of integrals over $z_0$, $z_1$ etc.), however, since the one-dimensional integral $\int\limits_0^{\infty}dz\,e^{-z}=1$, and only for finite amount of $z$-factors there appear nontrivial powers of $z$, the whole integral can be considered as effectively finite-dimensional. If, as before, among the $b$'s there appear elements $b_{m_1}$,  $b_{m_2}$ etc., up to some $b_{m_k}$, and they appear $\nu_1$ times, $\nu_2$ times etc, then the $z$ integral will eventually become
    \begin{equation}
        \int\limits_0^{\infty}\frac{Dz}{\nu!}\,z_{m_1}^{\nu_1}\ldots z_{m_k}^{\nu_k}\,e^{-z_0-z_1-\ldots}=\frac{\nu_1!\ldots \nu_k!}{\nu!}
    \end{equation}
    This is exactly the normalization factor that was needed. Therefore the eq. (\ref{AlgebraIntegralFull}) gives indeed a correctly defined inner product of the $(a,b)$ - algebra.
    \end{itemize}

     With this realization of the algebra we can finally rewrite the identity (\ref{DeltaWithAB}) as
    \begin{multline}\label{DeltaWithRIntegral}
    \int\frac{Dz}{\nu!}e^{-\sum\limits_r z_r}\int\limits_{(W)}dx\,\Biggl(\prod\limits_{i=1}^{\nu}\sum\limits_{m=0}^{\infty}f_m(\xi_i)z_m\,p_m^{-ix}\Biggr) \Biggl(\prod\limits_{j=1}^{\nu}\sum\limits_{n=0}^{\infty}f_n(\eta_j)\,p_n^{ix}\Biggr)=\\=\frac{1}{\nu!}\sum\limits_{P}\prod\limits_{i=1}^{\nu}\delta(\xi_{P(i)}-\eta_i)
\end{multline}

    Throughout the paper we will need to use a function $\delta^{(a,b)}(x-y)$, which plays a role of the Dirac $\delta$-function under $\int\limits_{(W)}$ integration. This function should satisfy the following identity for any function $F(x)$
    \begin{equation}\label{DeltaABDefinition}
         \int\limits_{(W)}dx\,\delta^{(a,b)}(x,y)\,F(x) = F(y)
    \end{equation}
    In order to construct the function $\delta^{(a,b)}(x,y)$ we use eq. (\ref{AlgebraIntegralCases}). Assume that $F(x)$ can be expanded into Dirichlet series
    \begin{equation}
        F(x)=\sum\limits_{n=0}^{\infty}\frac{f_n}{n^{ix}}
    \end{equation}
    Then it would be enough to establish for any integer $n\geq 0$
    \begin{equation}
        \int\limits_{(W)}dx\,e^{-i\,x\,\ln{n}}\,\delta^{(a,b)}(x,y) = e^{-i\,y\,\ln{n}}
    \end{equation}
    From eq. (\ref{AlgebraIntegralCases}) it follows that we can take
    \begin{equation}\label{ZetaDelta}
        \delta^{(a,b)}(x,y)=\sum\limits_{m=1}^{\infty}e^{i\, (x-y)\,\ln{m}}=\zeta\bigg(-i\,(x-y)\biggr)
    \end{equation}
Here $\zeta(x)$ is the Riemann zeta function.

\bibliographystyle{JHEP}
\bibliography{PathIntegralBibliography}

@article{PhysRevB.4.3174,
  title = {Renormalization Group and Critical Phenomena. I. Renormalization Group and the Kadanoff Scaling Picture},
  author = {Wilson, Kenneth G.},
  journal = {Phys. Rev. B},
  volume = {4},
  issue = {9},
  pages = {3174--3183},
  numpages = {0},
  year = {1971},
  month = {Nov},
  publisher = {American Physical Society},
  doi = {10.1103/PhysRevB.4.3174},
  url = {https://link.aps.org/doi/10.1103/PhysRevB.4.3174}
}

@article{PhysRevLett.28.240,
  title = {Critical Exponents in 3.99 Dimensions},
  author = {Wilson, Kenneth G. and Fisher, Michael E.},
  journal = {Phys. Rev. Lett.},
  volume = {28},
  issue = {4},
  pages = {240--243},
  numpages = {0},
  year = {1972},
  month = {Jan},
  publisher = {American Physical Society},
  doi = {10.1103/PhysRevLett.28.240},
  url = {https://link.aps.org/doi/10.1103/PhysRevLett.28.240}
}

@article{Banks:1981nn,
    author = "Banks, Tom and Zaks, A.",
    title = "{On the Phase Structure of Vector-Like Gauge Theories with Massless Fermions}",
    reportNumber = "TAUP-944-81",
    doi = "10.1016/0550-3213(82)90035-9",
    journal = "Nucl. Phys. B",
    volume = "196",
    pages = "189--204",
    year = "1982"
}

@article{Seiberg_1995,
   title={Electric-magnetic duality in supersymmetric non-Abelian gauge theories},
   volume={435},
   ISSN={0550-3213},
   url={http://dx.doi.org/10.1016/0550-3213(94)00023-8},
   DOI={10.1016/0550-3213(94)00023-8},
   number={1–2},
   journal={Nuclear Physics B},
   publisher={Elsevier BV},
   author={Seiberg, N.},
   year={1995},
   month=feb, pages={129–146} }

@article{Aharony_2000,
   title={Large N field theories, string theory and gravity},
   volume={323},
   ISSN={0370-1573},
   url={http://dx.doi.org/10.1016/S0370-1573(99)00083-6},
   DOI={10.1016/s0370-1573(99)00083-6},
   number={3–4},
   journal={Physics Reports},
   publisher={Elsevier BV},
   author={Aharony, Ofer and Gubser, Steven S. and Maldacena, Juan and Ooguri, Hirosi and Oz, Yaron},
   year={2000},
   month=jan, pages={183–386} }

@article{Comon2008,
  title = {Symmetric Tensors and Symmetric Tensor Rank},
  volume = {30},
  ISSN = {1095-7162},
  url = {http://dx.doi.org/10.1137/060661569},
  DOI = {10.1137/060661569},
  number = {3},
  journal = {SIAM Journal on Matrix Analysis and Applications},
  publisher = {Society for Industrial & Applied Mathematics (SIAM)},
  author = {Comon,  Pierre and Golub,  Gene and Lim,  Lek-Heng and Mourrain,  Bernard},
  year = {2008},
  month = jan,
  pages = {1254–1279}
}

\end{document}